\pdfoutput=1
\documentclass[fleqn,usenatbib]{mnras}

\usepackage{newtxtext,newtxmath}

\usepackage[T1]{fontenc}
\usepackage{ae,aecompl}

\usepackage{graphicx}	
\usepackage{amsmath}	
\usepackage{amssymb}	
\usepackage{gensymb}
\usepackage{hyperref}

\newcommand{\kms}{km s$^{-1}$}
\newcommand{\ms}{m s$^{-1}$}

\newcommand{\rjup}{R$_{\rm Jup}$}
\newcommand{\mjup}{M$_{\rm Jup}$}
\newcommand{\rhojup}{$\rho_{\rm Jup}$}
\newcommand{\mplanet}{M$_{\rm P}$}

\newcommand{\mstar}{M$_{\sun}$}
\newcommand{\rstar}{R$_{\sun}$}
\newcommand{\rhostar}{$\rho_{\sun}$}

\newcommand{\vsini}{$v\sin{i}$}

\newcommand{\KKstar}{EPIC~228735255}
\newcommand{\KKplanet}{EPIC~228735255b}

\newcommand{\KKRA}{$12:32:32.96$}
\newcommand{\KKDec}{$-09:36:27.5$}
\newcommand{\KKgaiaband}{$12.393$}
\newcommand{\KKBband}{$13.349\pm0.030$}
\newcommand{\KKVband}{$12.624\pm0.030$}
\newcommand{\KKgband}{$12.930\pm0.060$}
\newcommand{\KKrband}{$12.426\pm0.020$}
\newcommand{\KKiband}{$12.292\pm0.050$}
\newcommand{\KKJband}{$11.421\pm0.026$}
\newcommand{\KKHband}{$11.068\pm0.021$}
\newcommand{\KKKsband}{$10.995\pm0.021$}
\newcommand{\KKKepband}{$12.483$}
\newcommand{\KKWoneband}{$10.985\pm0.024$}
\newcommand{\KKWtwoband}{$11.030\pm0.021$}
\newcommand{\KKWthreeband}{$10.891\pm0.119$}
\newcommand{\KKWfourband}{$8.898\pm -$}
\newcommand{\KKdist}{$340.24\pm11.58$}
\newcommand{\KKage}{$4.22\pm0.95$}

\newcommand{\KKspectype}{$\rm G5$}
\newcommand{\KKvmag}{12.5}
\newcommand{\KKMv}{$4.965^{+0.069}_{-0.066}$}
\newcommand{\KKsmass}{$1.005^{+0.021}_{-0.020}$}
\newcommand{\KKsradius}{$0.987^{+0.011}_{-0.011}$}
\newcommand{\KKsrho}{$1.048\pm0.041$}
\newcommand{\KKslumin}{$0.893^{+0.049}_{-0.048}$}
\newcommand{\KKfeh}{$0.12\pm0.045$}
\newcommand{\KKsteff}{$5654\pm55$}
\newcommand{\KKslogg}{$4.452^{+0.010}_{-0.009}$}
\newcommand{\KKsvsini}{$3.8\pm0.2$}
\newcommand{\KKsrot}{$16.3\pm0.1$}
\newcommand{\KKsmuoneKtwo}{$0.341^{+0.084}_{-0.079}$}
\newcommand{\KKsmutwoKtwo}{$0.441^{+0.111}_{-0.079}$}
\newcommand{\KKsmuoneLCO}{$0.56^{+0.18}_{-0.16}$}
\newcommand{\KKsmutwoLCO}{$0.38^{+0.20}_{-0.17}$}
\newcommand{\KKsRMSC}{$0.0184$}
\newcommand{\KKsRMSH}{$0.0097$}

\newcommand{\KKperiod}{$6.569300^{+0.000017}_{-0.000020}$}
\newcommand{\KKshortperiod}{$\rm 6.57$}
\newcommand{\KKtc}{$2457588.28380^{+0.00014}_{-0.00014}$}
\newcommand{\KKpmass}{$1.019\pm0.070$}
\newcommand{\KKpradius}{$1.095\pm0.018$}
\newcommand{\KKprho}{$0.726\pm0.062$}
\newcommand{\KKplogg}{$3.324\pm0.033$}
\newcommand{\KKpeqtemp}{$1114\pm34$}
\newcommand{\KKK}{$0.1112^{+0.0076}_{-0.0073}$}
\newcommand{\KKe}{$0.120^{+0.056}_{-0.046}$}
\newcommand{\KKte}{$2.577$}
\newcommand{\KKw}{$98.88^{+3.85}_{-4.16}$}
\newcommand{\KKptdur}{$4.56\pm0.29$}
\newcommand{\KKptdurflat}{$3.49\pm0.26$}
\newcommand{\KKptgress}{$0.53\pm0.19$}
\newcommand{\KKpratio}{$0.1140^{+0.0015}_{-0.0012}$}
\newcommand{\KKpb}{$0.33\pm0.14$}
\newcommand{\KKpinc}{$88.51^{+0.69}_{-0.53}$}
\newcommand{\KKpa}{$0.0591\pm0.0034$}
\newcommand{\KKpgammaC}{$1.2170^{+0.0089}_{-0.0092}$}
\newcommand{\KKpgammaH}{$1.2435^{+0.0068}_{-0.0070}$}
\newcommand{\KKpjitterC}{$0.0041^{+0.0127}_{-0.0037}$}
\newcommand{\KKpjitterH}{$0.0037^{+0.0144}_{-0.0033}$}
\newcommand{\KKpicflux}{$2.565\pm0.105$}

\title[\KKplanet]{\KKplanet{} -- An eccentric \KKshortperiod\,day transiting hot Jupiter in Virgo}

\author[H.A.C. Giles et al.]{
H.A.C. Giles,$^{1}$\thanks{E-mail: Helen.Giles@unige.ch}
D. Bayliss,$^{1}$
N. Espinoza,$^{2,3}$
R. Brahm,$^{2,3}$
S. Blanco-Cuaresma,$^{1,4}$
\newauthor
A. Shporer,$^{5}$ 
D. Armstrong,$^{6}$
C. Lovis,$^{1}$
S. Udry,$^{1}$
F. Bouchy,$^{1}$
M. Marmier,$^{1}$
\newauthor
A. Jord\'{a}n,$^{2,3}$
J. Bento,$^{7}$
A. Collier~Cameron,$^{8}$
R. Sefako,$^{9}$
W. D. Cochran,$^{10}$
\newauthor
F. Rojas,$^{2}$
M. Rabus,$^{2}$
J.S. Jenkins,$^{11}$
M. Jones,$^{12}$
B. Pantoja,$^{11,12}$
M. Soto,$^{11}$
\newauthor
R. Jensen-Clem,$^{13}$
D.A. Duev,$^{13}$
M. Salama,$^{14}$
R. Riddle,$^{13}$
C. Baranec,$^{14}$
N.M. Law$^{15}$
\\
$^{1}$Observatoire de Gen\`{e}ve, Universit\'{e} de Gen\`{e}ve, Chemin des Maillettes 51, 1290 Versoix, Switzerland\\
$^{2}$Instituto de Astrof\'isica, Facultad de F\'isica, Pontificia Universidad Cat\'{o}lica de Chile, Av. Vicu\~na Mackenna 4860, \\
782-0436 Macul, Santiago, Chile.\\
$^{3}$Millennium Institute of Astrophysics, Av. Vicu\~na Mackenna 4860, 782-0436 Macul, Santiago, Chile.\\
$^{4}$Harvard-Smithsonian Center for Astrophysics, 60 Garden Street, Cambridge, MA 02138, USA\\
$^{5}$Division of Geological and Planetary Sciences, California Institute of Technology, Pasadena, CA 91125, USA\\
$^{6}$Department of Physics, University of Warwick, Gibbet Hill Road, Coventry, CV4 7AL, UK\\
$^{7}$Research School of Astronomy and Astrophysics, Mount Stromlo Observatory, Australian National University, Weston,\\ ACT 2611, Australia.\\
$^{8}$Centre for Exoplanet Science, SUPA School of Physics \&\ Astronomy, University of St Andrews, \\
North Haugh, St Andrews KY16 9SS, UK\\
$^{9}$South African Astronomical Observatory, PO Box 9, Observatory, 7935.\\
$^{10}$McDonald Observatory and Department of Astronomy, University of Texas at Austin, TX, USA\\
$^{11}$ Departamento de Astronom\'ia, Universidad de Chile, Casilla 36-D, Santiago, Chile.\\
$^{12}$European Southern Observatory, Alonso de Cordova 3107, Vitacura, Casilla 19001, Santiago, Chile.\\
$^{13}$Department of Astronomy, California Institute of Technology, 1200 E. California Blvd., Pasadena, CA 91101, USA\\
$^{14}$Institute for Astronomy, University of Hawai`i at M\={a}noa, Hilo, HI 96720-2700, USA.\\
$^{15}$Department of Physics and Astronomy, University of North Carolina at Chapel Hill, Chapel Hill, NC 27599-3255, USA.\\
}

\date{Accepted XXX. Received YYY; in original form ZZZ}

\pubyear{2017}

\begin{document}
\label{firstpage}
 
\pagerange{\pageref{firstpage}--\pageref{lastpage}}
\maketitle

\begin{abstract}
We present the discovery of \KKplanet, a P=\KKshortperiod\,days Jupiter-mass (\mplanet=\KKpmass\mjup) planet transiting a V=\KKvmag\ (\KKspectype-spectral type) star in an eccentric orbit (e=\KKe) detected using a combination of \textit{K2} photometry and ground-based observations.  With a radius of \KKpradius\,\rjup, the planet has a bulk density of \KKprho\,\rhojup. The host star has a [Fe/H] of \KKfeh, and from the \textit{K2} light curve we find a rotation period for the star of \KKsrot\, days. This discovery is the 9th hot Jupiter from \textit{K2} and highlights \textit{K2}'s ability to detect transiting giant planets at periods slightly longer than traditional, ground-based surveys.
This planet is slightly inflated, but much less than others with similar incident fluxes. These are of interest for investigating the inflation mechanism of hot Jupiters.

\end{abstract}




\section{Introduction}

Transiting exoplanets offer the best insight into worlds outside our Solar System, as we can determine the mass, radius, and obtain information regarding the planetary atmosphere. 
Traditional ground-based surveys such as HAT-Net \citep{bakos:2004:hatnet}, WASP \citep{2006PASP..118.1407P} and KELT \citep{2007PASP..119..923P} are predominately sensitive to very short period transiting giant planets (P$\sim3$d).  Longer period transiting systems have proved much more difficult to detect.  Some advantage has been gained using multi-site surveys, with HATSouth \citep{bakos:2013:hatsouth} detecting planets in periods as long as 16 days \citep{HATS17}.
However, the continuous monitoring enabled by space-based telescopes has allowed for a dramatic increase in the number of longer period transiting systems. The \textit{Kepler} mission \citep{Borucki_et_al_2010,Koch_et_al_2010,Jenkins_et_al_2010}, with 4 years of near-continuous coverage, has uncovered a host of transiting planets with longer periods, however many of these transit stars that are too faint to allow for planetary mass determination via radial velocities.  In 2013, after 4 years of observations, the second of \textit{Kepler}'s four reaction wheels failed. From this, the \textit{K2} mission was born \citep{Howell2014}. Unlike the original mission, which observed a single region of the sky, \textit{K2} observes proposed targets within a series of fields lying along the ecliptic continuously for $\sim80$ days. The adverse impact of the two failed reaction wheels has been minimised, but there is now a 6-hour roll effect affecting \textit{K2} light curves. This causes brightness changes as stars move from pixel to pixel on the CCD. However, there have been many different attempts to calibrate this effect and remove it from the light curves allowing for transiting exoplanet searches \citep{K2SFF,K2VARCAT,K2SC}.
Additionally, the continuous observations for 80 days still allows for longer period systems to be discovered, e.g. EPIC~201702477b \citep[40.736d,][]{Bayliss2017}.  Additionally a number of more typical hot Jupiters have been discovered, e.g. K2-30b (4.099d), K2-34b (2.996d) \citep{K2_30b_34b} and K2-31b \citep[1.258d,][]{K2_31b}.

In this paper we report the discovery of \KKplanet, a \KKshortperiod\, day hot Jupiter. In \S2 we outline the observations that led to the discovery. In \S3 we describe the analysis of the data which determined its properties. In \S4, we discuss the properties and the planet's position with respect to other known hot Jupiters, and in \S5 we summarise the discovery.

\section{Observations}

In this section we set out the observations made to detect and characterise the transiting exoplanet \KKplanet.

\begin{table*}
	\centering
	\caption{Parameters of \KKstar}
	\label{tab:starparams}
	\begin{tabular}{lllr}
		\hline
        \hline
		Parameter & Units & Value & Source\\
        \hline
        EPIC ID & & 228735255 & H16*\\
        2MASS ID & & 2MASS J12323296-0936274 & H16*\\
        R.A. ($\rm\alpha$) & hh:mm:ss& \KKRA & H16*\\
        Dec. ($\rm\delta$) & dd:mm:ss& \KKDec & H16*\\
        \\
        g$_{\rm GAIA}$ & mag & \KKgaiaband & \textsc{GAIA}$^\circ$\\
        B & mag & \KKBband & \textsc{APASS}$^\star$ \\
        V & mag & \KKVband & \textsc{APASS}$^\star$ \\
        g & mag & \KKgband & \textsc{APASS}$^\star$ \\
        r & mag & \KKrband & \textsc{APASS}$^\star$ \\
        i & mag & \KKiband & \textsc{APASS}$^\star$ \\
        J & mag & \KKJband & \textsc{APASS}$^\star$ \\
        H & mag & \KKHband & \textsc{APASS}$^\star$ \\
        K$_{\rm s}$ & mag & \KKKsband & \textsc{2MASS}$^\bullet$\\
        Kep & mag & \KKKepband & H16*\\
        W$_{\rm 1}$ & mag & \KKWoneband & \textsc{AllWISE}$^+$\\
        W$_{\rm 2}$ & mag & \KKWtwoband & \textsc{AllWISE}$^+$\\
        W$_{\rm 3}$ & mag & \KKWthreeband & \textsc{AllWISE$^+$}\\
        W$_{\rm 4}$ & mag & \KKWfourband & \textsc{AllWISE}$^+$\\
        \\
        Distance & pc & \KKdist & $^\dagger$\\
        Age & Gyr & \KKage & $^\dagger$\\
        Spectral Type & & \KKspectype & $^\dagger$\\
        M$_{\rm V}$ & mag & \KKMv & $^\dagger$\\
        $[\rm Fe/H]$ & dex & \KKfeh & $^\dagger$\\
        T$_{\rm eff}$ & K & \KKsteff & $^\dagger$\\
        log(g) & dex & \KKslogg & $^\dagger$\\
        \textit{v}sin\textit{i} & \kms & \KKsvsini & $^\dagger$\\
        P$_{\rm rot}$ & days & \KKsrot & $^\dagger$\\
        M$_*$ & \mstar & \KKsmass & $^\dagger$\\
        R$_*$ & \rstar & \KKsradius & $^\dagger$\\
        $\rho_*$ & \rhostar & \KKsrho & $^\dagger$\\
        $\rm L_*$ & $L_{\sun}$ & \KKslumin & $^\dagger$\\
        $\mu_{1,\rm K2}$ & & \KKsmuoneKtwo & $^\dagger$\\
        $\mu_{2,\rm K2}$ & & \KKsmutwoKtwo & $^\dagger$\\
        $\mu_{1,\rm LCO}$ & & \KKsmuoneLCO & $^\dagger$\\
        $\mu_{2,\rm LCO}$ & & \KKsmutwoLCO & $^\dagger$\\
        RV residuals (CORALIE) & \kms & \KKsRMSC & $^\dagger$\\
        RV residuals (HARPS) & \kms & \KKsRMSH & $^\dagger$\\
        \hline
        \hline
        \multicolumn{3}{l}{* \citet{Huber2016}, $^\circ$ \citet{GAIAa,GAIAb},}\\
        \multicolumn{3}{l}{$^\star$ \citet{APASS},$^\bullet$ \citet{2MASS},}\\
        \multicolumn{3}{l}{$^+$ \citet{WISE,NEOWISE}, $^\dagger$ This work}\\
	\end{tabular}
\end{table*}

\subsection{\textit{K2} Photometry}
\label{sec:K2}

\begin{table}
	\centering
	\caption{Photometry for \KKstar}
	\label{tab:k2lc}
	\begin{tabular}{llllr}
		\hline
        \hline
		BJD-2450000 & Flux & Flux Error & Filter & Instr.\\
		\hline
        7582.5906314203& 1.00002094& 0.00008399& kep& K2\\
        7582.6110636177& 1.00006507& 0.00008393& kep& K2\\
        7582.6314957142& 1.00000389& 0.00008385& kep& K2\\
        7582.6519277110& 1.00001766& 0.00008376& kep& K2\\
        7582.6723599071& 1.00000462& 0.00008369& kep& K2\\
        7582.6927920026& 0.99998520& 0.00008362& kep& K2\\
        7582.7132239980& 1.00001908& 0.00008351& kep& K2\\
        7582.7336561931& 0.99992969& 0.00008342& kep& K2\\
        7582.7540882872& 1.00015249& 0.00008331& kep& K2\\
        7582.7745202812& 0.99993307& 0.00008323& kep& K2\\
        ...& ...& ...& ...& ...\\
		\hline
        \hline
        \multicolumn{5}{l}{* Note: partial list -- full table available in electronic form.}\\
	\end{tabular}
\end{table}

The light curve for \KKstar\, came from Campaign 10 of the \textit{K2} mission. This campaign observed 41607 targets in Long Cadence (30 minutes) and 138 in Short Cadence (1 minute) in the ecliptic plane centered around RA 12h 27m 07.07s Dec -04$\degree\,$01'\,37.77''. Due to a pointing error (targets were off by 12 arcseconds meaning many fell outside their apertures) this campaign was split into two data releases, C10a and C10b. C10a lasted 6 days between 2016 July 6 19:45:29 UTC and 2016 July 13 01:19:55 UTC. The second release, C10b, was observed for 69 days. However there was a data gap of 14 days after 7 days of observing due to module 4 of the telescope failing which powered off the photometer.

\begin{figure*}
	\includegraphics[width=\textwidth]{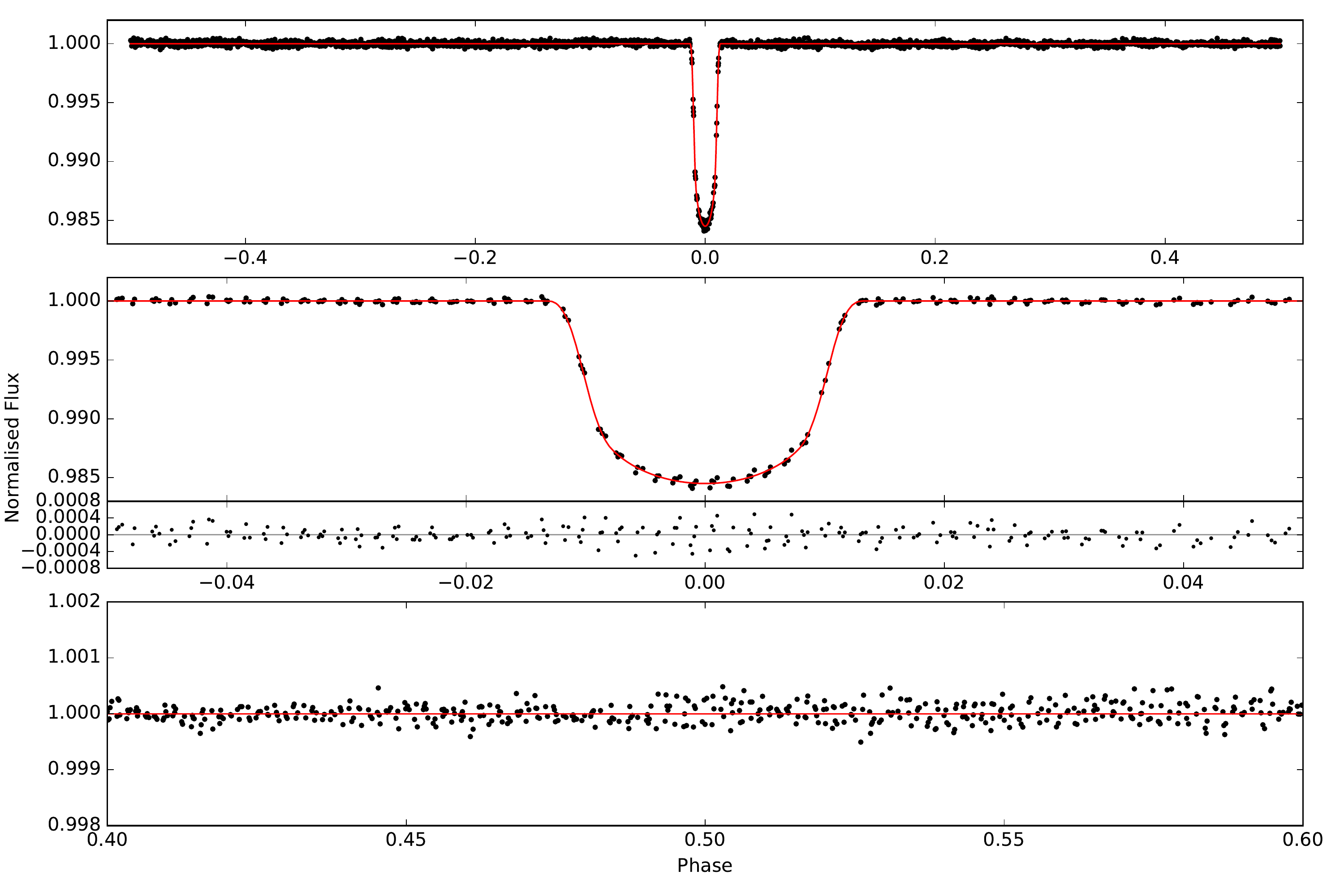}
    \caption{Phase-folded \textit{K2} light curve of \KKstar\, (black points) with best-fit model plotted as a solid red line (see \S~\ref{sec:jointfit}). Top panel: Full phase light curve with the transit of \KKplanet. There are no other significant dips indicating any other transits. Middle panel: Zoom-in of the transit of \KKplanet\, and the resulting residuals from it and the model fit. Bottom panel: Zoom-in around phase 0.5.  There is no indication of an observable secondary eclipse.}
    \label{fig:k2_transit}
\end{figure*}

After the public release of the data on 2016 December 20, the light curves reduced by the \textit{K2} Science team were downloaded and analysed for planetary signals (light curve data listed in Table~\ref{tab:k2lc}).
This analysis required long term variations to be removed from the light curve. This was done by fitting a sliding polynomial, which fits locally a polynomial to a small section (`stepsize') of the light curve using a significantly larger section (`window') of the surrounding light curve, and dividing it out. For the sliding polynomial, we used a 3rd order polynomial with a stepsize of 0.1 days and a window size of 5 days. 
To ensure the result is not jagged, the stepsize must be significantly smaller than the window size and to ensure that the transit is not accidentally fitted and removed by the process, to ensure the transit is left intact requires outlier rejection from the polynomial fit -- this was done with a strict cut of positive outliers and a looser negative outlier cut.
To search for planetary transits, we used a python-wrapped\footnote{\url{https://github.com/dfm/python-bls}} version of the BLS routine \citep{kovacs:2002:BLS} to initially search for any significant signals and then a second time focused on the signal of interest to determine the transit parameters as accurately as possible. We then phase fold and output the light curve for visual inspection.
This transit search found many candidates, which included \KKstar\,-- a \KKshortperiod\, day planet with a 1.26 per cent transit signal (Fig.~\ref{fig:k2_transit}).

Additionally, as can be seen in Fig.~\ref{fig:k2_transit}, there is some evidence of aliasing in the cadence. This is due to the observed rotation period being a half integer multiple of the cadence of \textit{K2}.

\subsection{Radial Velocities}
\label{sec:rv}
We observed \KKstar\ using the CORALIE spectrograph \citep{Queloz2000} on the 1.2\,m Euler Telescope at La Silla Observatory in Chile. CORALIE is a fibre-fed, high resolution (R=60,000) echelle spectrograph capable of delivering <6 \ms{} accuracy. Observations were made between 2017 February 20 and 2017 April 8.
Additionally, \KKstar\ was observed using the High Accuracy Radial Velocity
Planet Searcher \citep[HARPS,][]{MayorHARPS:2003} mounted on the ESO 3.6m telescope in La Silla Observatory in Chile, on February 22 and between April 23 and 28. The spectra, which have a resolution R = 115000, were reduced using the Collection of Elemental Routines for Echelle Spectra \citep[CERES,][]{CERES:2017}.

The associated errors with each instrument vary significantly. In the case of CORALIE, the initial errors are higher than HARPS primarily because the star is relatively faint.  As a test for the errors, we also calculated the root-mean-square of the data points from the fitted model (see \S~\ref{sec:jointfit}) and they were comparable to the measured errors (see Tab.~\ref{tab:starparams}).

\begin{figure}
	\includegraphics[width=\columnwidth]{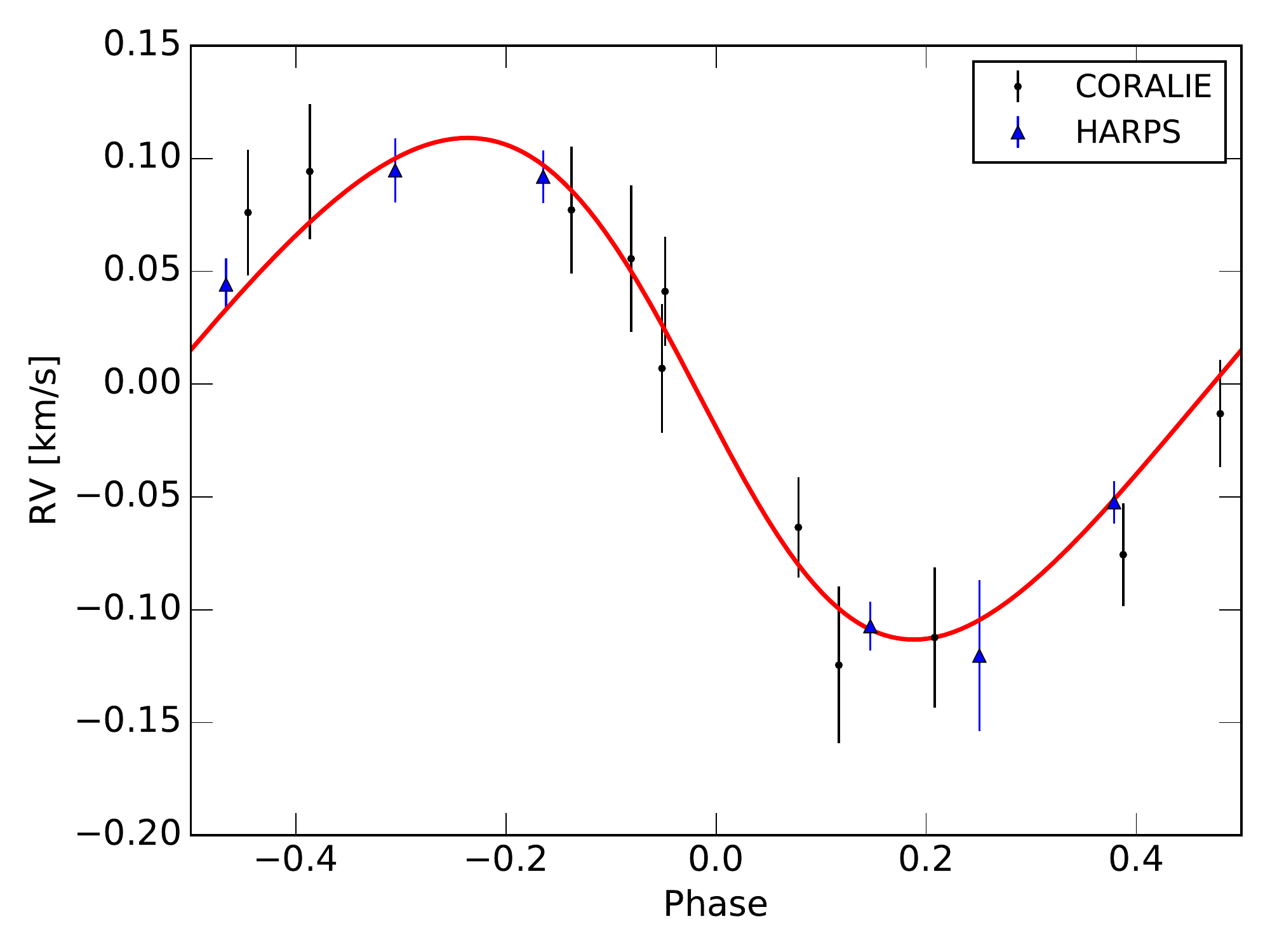}
    \caption{Phase-folded observations from CORALIE (black points) and HARPS (blue triangles) with the best-fit model (red solid line) as described in \S~\ref{sec:jointfit}. CORALIE observations were taken between 2017 February 20 and 2017 April 8 using the Swiss Euler telescope in La Silla, Chile. HARPS observations were taken between 2017 February 22 and 2017 April 28 using the ESO 3.6m telescope in La Silla, Chile. The errors for HARPS have had the jitter added in quadrature.}
    \label{fig:rv_curve}
\end{figure}

The radial velocities are plotted in Fig.~\ref{fig:rv_curve}, along with the best fit model determined by the joint fit described in \S~\ref{sec:jointfit}.  The radial velocities are also presented in Table~\ref{tab:rv}.

\begin{table}
	\centering
	\caption{Radial Velocities for \KKplanet\, in chronological order}
	\label{tab:rv}
	\begin{tabular}{llllr}
		\hline
        \hline
		BJD-2450000 & RV & RV error & BIS & Instrument\\
         & \kms & \kms & & \\
		\hline
        7804.751722&	1.25803&	0.02389&	-0.03886&  CORALIE\\	
        7806.7167899&   1.1231&     0.0333&     -0.019&    HARPS\\
        7814.792453&	1.20384&	0.02344&	-0.08219&  CORALIE\\
		7815.668909&	1.31119&	0.02967&	-0.03487&  CORALIE\\
		7817.678137&	1.27255&	0.03227&	-0.06864&  CORALIE\\
		7818.723682&	1.15348&	0.02183&	-0.01804&  CORALIE\\
		7820.754834&	1.14137&	0.02247&	0.02071&   CORALIE\\
		7821.852054&	1.29300&	0.02745&	0.03278&   CORALIE\\
		7823.874191&	1.29414&	0.02778&	0.02189&   CORALIE\\
		7832.714263&	1.10465&	0.03093&	-0.06040&  CORALIE\\
		7836.681188&	1.31122&	0.03383&	0.00470&   CORALIE\\
        7850.717632&	1.22394&	0.02820&	-0.05072&  CORALIE\\
        7851.823385&	1.09243&	0.03450&	-0.00820&  CORALIE\\
        7866.6826602&   1.1911&     0.0086&     0.02&      HARPS\\
        7867.699734&    1.2876&     0.0111&     -0.009&    HARPS\\
        7868.7574426&   1.3382&     0.0137&     0.003&     HARPS\\
        7869.6828912&   1.3354&     0.0111&     0.039&     HARPS\\
        7871.7285204&   1.1362&     0.0102&     0.0&       HARPS\\
		\hline
        \hline
	\end{tabular}
\end{table}

\begin{figure}
	\includegraphics[width=\columnwidth]{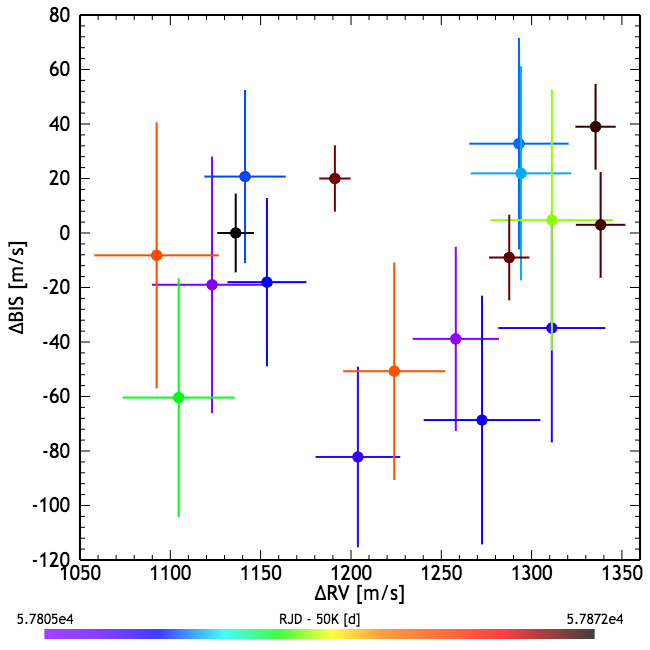}
    \caption{Distribution of the measured radial velocities and associated bisector slopes from CORALIE and HARPS. No evidence of correlation between the two can be seen. Colours represent date of observation between 2017 February 20 and 2017 April 28.}
    \label{fig:bisector}
\end{figure}

In order to check radial velocity variation induced by a blended spectrum, we computed the bisector slope of the cross-correlation function for each observation in the manner described in \citep{Queloz2001}. In Fig.~\ref{fig:bisector} we find no correlation between the bisector slope and the measured radial velocity. If the signal detected was due to a blended eclipsing binary, then we may expect to see a strong correlation between the bisectors and radial velocity measurements.  The bisector values are presented with the radial velocities in Table~\ref{tab:rv}.

\subsection{LCO photometry}
\label{sec:lco}
\begin{figure}
	\includegraphics[width=\columnwidth]{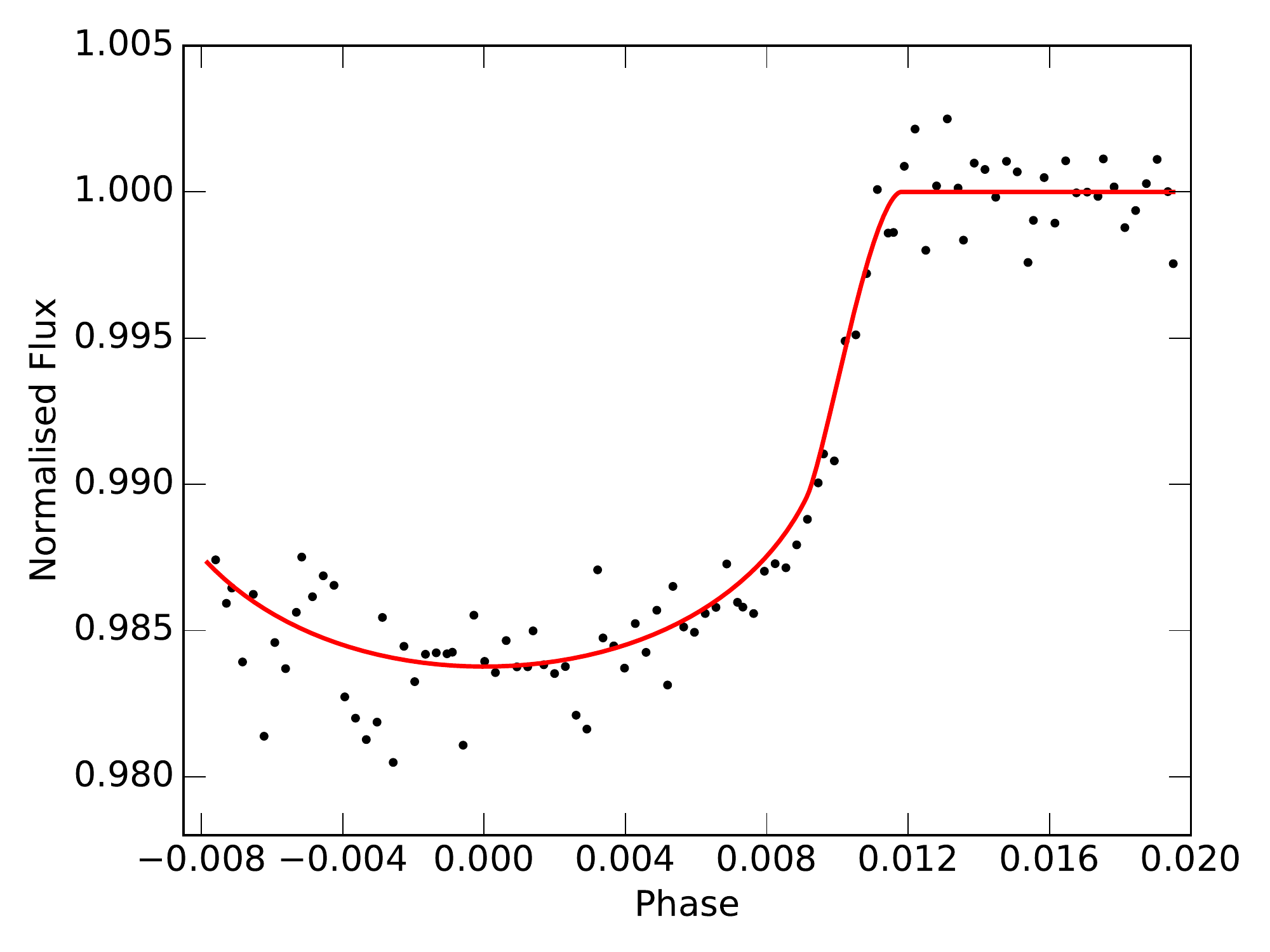}
    \caption{Phase-folded LCO light curve of \KKstar\, (black points) with best-fit model plotted as a solid red line (see \S~\ref{sec:jointfit}). Observations occurred the night of 2017 March 18 at the 1m LCO telescope in Sutherland, South Africa, in `i' band.}
    \label{fig:lco_transit}
\end{figure}

In order to refine the ephemeris, check for TTVs, and check for a colour-dependent transit depth (signifying a probable blend), we performed ground-based photometric follow-up using the Las Cumbres Observatory (LCO) 1-m telescope network \citep{brown:2013:lcogt}. On 2017 March 18 we monitored the transit in the i-band using the three LCO 1-m telescopes situated at South Africa Astronomical Observatory at Sutherland, South Africa (Fig.~\ref{fig:lco_transit}). The observations were taken using the ``Sinistro'' camera with exposure times of 120\,s and the telescope defocused (2.0mm) to avoid saturation and spread the stellar point-spread function over more pixels -- reducing the impact of flat-fielding uncertainties. The images were reduced using the standard LCO reduction pipeline (BANZAI), and then aperture photometry was performed using an automated pipeline (Espinoza et al., 2017, in prep). These observations were made as part of a wider LCO Key Project\footnote{\url{http://web.gps.caltech.edu/~shporer/LCOKP/}} to characterize transiting planets using the LCO 1-m network \citep[see][]{Bayliss2017}. They are listed in Table~\ref{tab:k2lc}.

\subsection{High Angular Resolution Imaging}
\label{sec:ao}
High angular resolution imaging of the target was obtained using the Robo-AO instrument \citep{baranec13, baranec14} mounted on the Kitt Peak 2.1m telescope, on the night of 2017 April 15 using the long-pass ``lp600" filter \citep{baranec14} with a seeing of 1.5" with a resulting Strehl ration of 2.7\%. The raw rapid read-out data from the Robo-AO visual camera were processed using Robo-AO's reduction pipelines described briefly below. A more detailed description can be found in \citet{jensen-clem17}.

First, the ``bright star" pipeline generates a windowed data cube centered on an automatically selected guide star. The windowed region is bi-cubically up-sampled and cross correlated with the theoretical PSF to give the center coordinates of the guide star's PSF in each frame. The nightly dark and dome flat exposures are then used to calibrate the full-frame, unprocessed images. The calibrated full frames are aligned using the center coordinates identified by the up-sampled, windowed frames, and co-added via the Drizzle algorithm.

Next, the ``high contrast imaging pipeline" generates a 3.5'' frame windowed about the star of interest in the final science frame from the bright star pipeline. A high pass filter is applied to the windowed frame to reduce the contribution of the stellar halo. To whiten correlated speckle noise at small angular separations from the target star, a synthetic PSF generated by the Karhunen-Lo\`{e}ve Image Processing (KLIP) algorithm is subtracted from the frame. The KLIP algorithm is based on the method of Principal Component Analysis. The PSF diversity needed to create this synthetic image is provided by a reference library of Robo-AO observations --- a technique called Reference star Differential Imaging.

The contrast curve was estimated using the \texttt{Vortex Image Processing (VIP)} package \citep{gomez_gonzalez_VIP} by measuring the residuals from resolution element-sized regions in the PSF-subtracted image.

\begin{figure}
	\includegraphics[width=\columnwidth]{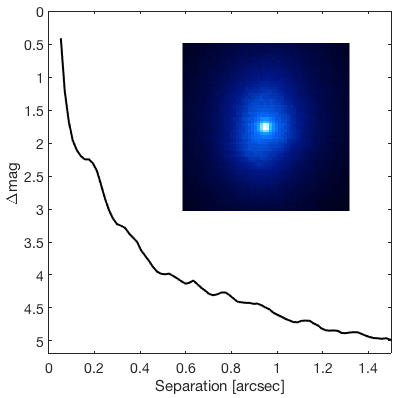}
    \caption{Contrast curve of \KKstar\ showing the upper limit on the magnitude difference between the target and a possible nearby star as a function of angular separation in arcsec. Data was taken by Robo-AO with the long-pass filter lp600 covering a wavelength range from 600 nm to close to 1 \micron\ \citep{baranec14, jensen-clem17}. The inset shows the image of the target spanning 1 arcsec on the side.}
    \label{fig:contrast_curve}
\end{figure}

The final Robo-AO image and contrast curve are shown in Fig.~\ref{fig:contrast_curve}. The target is isolated down to $\Delta$mag = 4 at 0.5 arcsec and $\Delta$mag = 4.5 at 1 arcsec.

\section{Analysis}
\label{sec:analysis}

\subsection{Stellar Parameters}
\label{sec:starparam}
Initially, to determine the stellar parameters of \KKstar\, we built a pipeline for CORALIE spectra based on \textsc{iSpec}\footnote{\url{http://www.blancocuaresma.com/s/iSpec}} \citep{ispec}. This tool provides a large number of options to treat high-resolution spectra (e.g., co-addition, continuum normalization) and it can derive atmospheric parameters and chemical abundances using many different model atmospheres, atomic line lists, radiative transfer codes and spectroscopic techniques (i.e., equivalent width and synthetic spectral fitting). For this study, we executed the following steps:

\begin{itemize}
	\item Align and co-add all the observations taken with CORALIE (see Section~\ref{sec:rv}) to increase the S/N.
    \item Reduce the spectrum to the optical wavelength range (480 - 680 nm).
    \item Cross-correlate with a solar template to shift the observed spectrum to the rest frame.
    \item Discard negative fluxes and estimate flux errors based on an estimated S/N.
    \item Convolve to a resolution of $R\sim47\,000$ and homogeneously re-sample the spectrum.
    \item Ignore regions affected by telluric lines.
    \item Fit the pseudo-continuum and normalize the spectrum.
    \item Derive atmospheric parameters using the synthetic spectral fitting technique, SPECTRUM \citep{1994AJ....107..742G} as radiative transfer code, atomic data obtained from VALD \citep{2011BaltA..20..503K}, a line selection based on a $R\sim47\,000$ solar spectrum \citep{2016csss.confE..22B, 2017hsa9.conf..334B} and the MARCS model atmospheres \citep{MARCS_model}.
\end{itemize}

As an output we obtained the effective temperature, surface gravity (log g) and metallicity, which is basically correlated with the iron abundance (i.e., [Fe/H]) of the star. From these, a series of isochrones were generated using stellar model generator \textsc{SYCLIST}\footnote{\url{https://obswww.unige.ch/Recherche/evoldb/index/}} \citep{syclist}. A grid of ages at a given metallicity (Z=0.040) were generated and interpolated to determine the stellar age, mass, radius and luminosity.

The results of the \textsc{iSpec} analysis gave an effective temperature of 5732$\pm$32K, a log g of 4.29 dex and [Fe/H] = 0.32$\pm$0.03 dex.

Following a similar procedure, the individual HARPS spectra were median combined in order to construct a higher SNR template. The resulting spectrum was used as input of the Zonal Atmospheric Parameter estimator \citep[ZASPE,][]{zaspe} for computing the stellar atmospheric parameters (T$_{\rm eff}$, log$g$, [Fe/H] and v$_{\rm rot}\sin{i}$) by comparing it with a grid of synthetic spectra generated from the ATLAS9 model atmospheres \citep{kurucz}.

For estimating an initial guess of the physical parameters of the star we used the Yonsei-Yale Isochrones \citep{yy} by searching for the $M_{*}$ and stellar Age of the model that would produce the observed T$_{\rm eff}$ and $a/R_{*}$ values for the given [Fe/H]. For obtaining the errors in the physical parameters we performed Monte Carlo simulations where new values for T$_{\rm eff}$, $a/R_{*}$ and [Fe/H] were sampled from Gaussian distributions in each realisation.

\begin{figure}
	\includegraphics[width=\columnwidth]{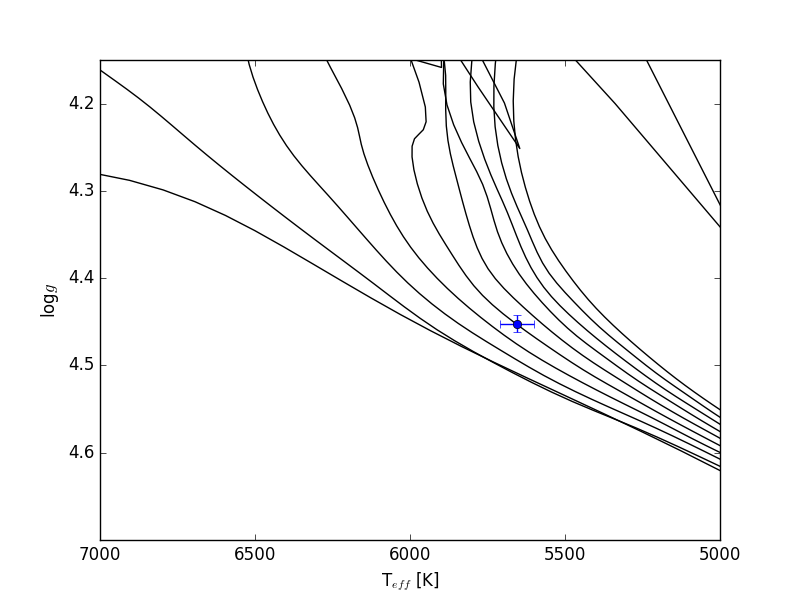}
    \caption{Yonsei-Yale isochrones \citep{yy} covering (from left to right) 0.1, 1, 2, 3, 4, 5, 6, 7, 8 and 9 Gyr based on the stellar parameters determined for \KKstar.}
    \label{fig:isochrones}
\end{figure}

The resulting physical parameters were used to compute a more precise value for the stellar log$g$ than the one obtained from spectroscopy. The new log$g$ value was then held fixed in a new ZASPE execution whose results are displayed in Table 1. The new atmospheric parameters were used to determine a new set of physical parameters from the Yonsei-Yale Isochrones (Fig.~\ref{fig:isochrones}) obtaining a stellar mass of 1.005 $\pm$ 0.020 M$_{\odot}$, a stellar age of 4.2 $\pm$ 1.0 Gyr and a corresponding stellar radius of 0.987 $\pm$ 0.011 R$_{\odot}$, making this host star a slightly metal-rich solar analogue.

Comparing the two results (\textsc{iSpec} and ZASPE), a very similar effective temperature was measured. However, the \textsc{iSpec} routine detected a smaller log g and higher metallicity. We attribute this discrepancy to the HARPS data having a significantly better signal-to-noise ratio than the reconnaissance spectroscopy which was measured from CORALIE.
For further analysis, the results from ZASPE using HARPS data were used.

\begin{figure}
	\includegraphics[width=\columnwidth]{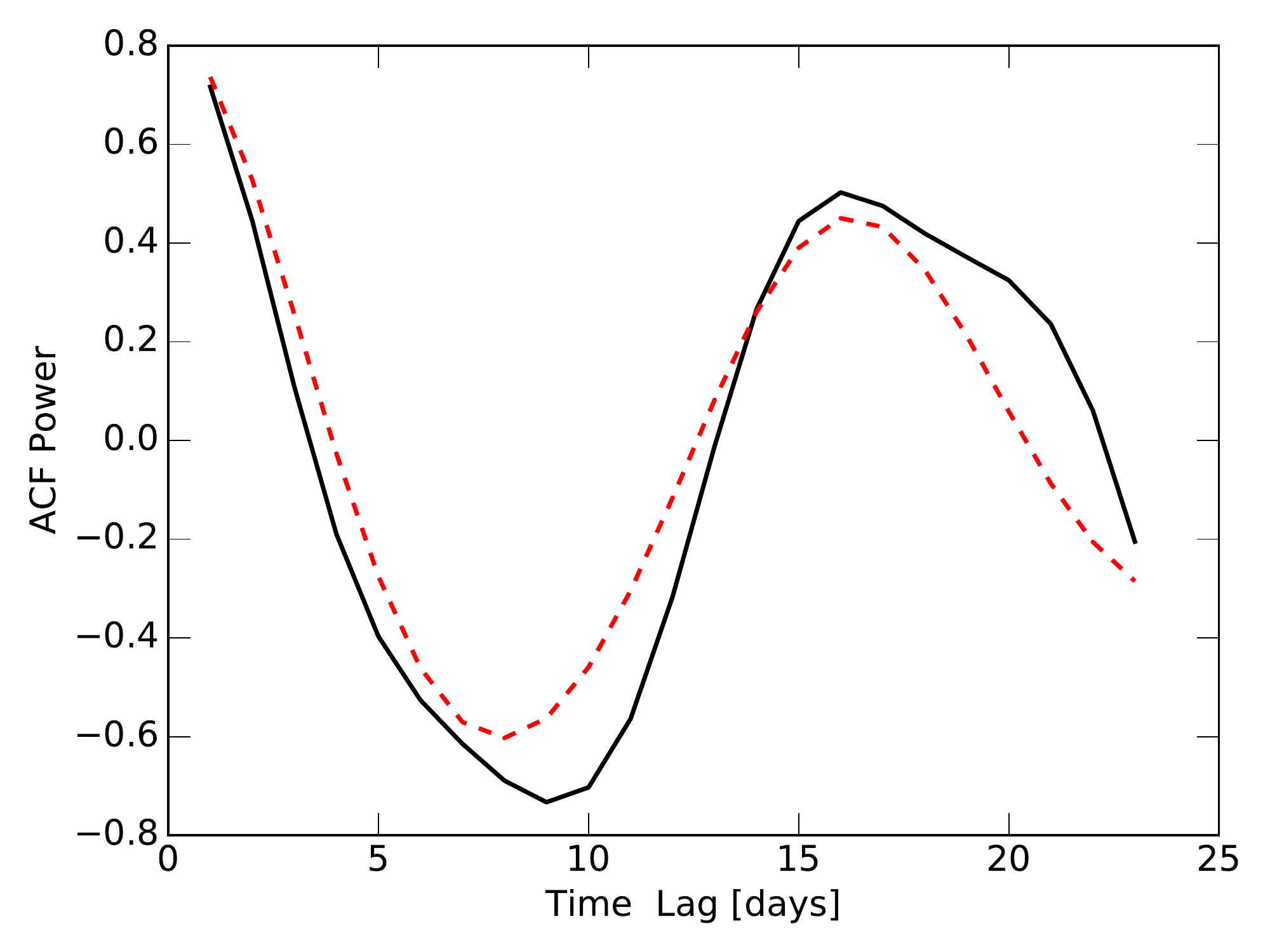}
    \caption{The autocorrelation function (black line) of the light curve for \KKstar\, with the transits omitted, fitted with a harmonic function (red dashed line) using method described in \citet{Giles17}. This measured a rotation period for \KKstar\, of \KKsrot\, days.}
    \label{fig:acf}
\end{figure}

We measured the rotation period of \KKstar\, using an auto-correlation function of the \textit{K2} light curve (with the transit omitted) as described in \citep{Giles17}. This determined a rotation period of \KKsrot\, days (Fig.~\ref{fig:acf}).  Given R$_*$ = \KKsradius, this rotation should result in a \vsini $\sim 3.1$\,\kms assuming stellar spin axis is perpendicular to the orbital plane of the planet.  The spectroscopically derived \vsini\, is slightly larger than this value (\vsini=\KKsvsini), which may be due to non-equatorial spots and solar-like differential rotation.  Such an effect has been seen in other \textit{K2} transiting systems, e.g. HATS-36b \citep{Bayliss17b}.

\subsection{Joint Fit}
\label{sec:jointfit}
We fit the photometry data from \S~\ref{sec:K2} and \S~\ref{sec:lco} jointly with the radial velocities from \S~\ref{sec:rv} using the \texttt{exonailer}\footnote{\url{https://github.com/nespinoza/exonailer}} algorithm \citep{EXONAILER}. The \texttt{exonailer} fitting was conducted with loose priors on the period, P, time of first transit, T$_0$, and planet-to-stellar-radii ratio, $p$ (see Table~\ref{tab:planetparams} for priors used). These were determined directly from the \textit{K2} light curve. In addition, extra (Gaussian) noise terms were added to the errors of the LCO and \textit{K2} photometry (in order to empirically estimate extra photometric jitter), with a prior of $\mathcal{N}(1,1000)$ for each.  Extra Gaussian noise terms were also added to the CORALIE and HARPS radial velocities (in order to model radial-velocity jitter either instrumental or from stellar origin due to, e.g., activity).

Special care was taken in the modelling of the limb-darkening effect, as it is known that this can have a direct impact on the retrieved fitted transit parameters \citep{EJ:2015}. In order to select the best limb-darkening law, we followed \cite{EJ:2016} and ran the \texttt{ld-exosim} algorithm\footnote{\url{https://github.com/nespinoza/ld-exosim}}, which gives the mean-square error on each of the retrieved transit parameters for a given limb-darkening law (given the noise, sampling and geometry of the transit). The quadratic law was chosen as it was the law that gave the minimum mean-square error on the planet-to-star radius ratio. For this case, this was the most important transit parameter because it defines the exoplanet's density. Additionally, the limb-darkening coefficients were individually fitted for the \textit{K2} and LCO light curves, as they have different response functions and span different wavelength ranges. An initial fit assuming different planet-to-star radius ratios for each dataset was also made, but both independently gave consistent parameters with no wavelength dependence. The final fit was made by using a common planet-to-star radius ratio for both datasets.  Priors for the limb-darkening coefficients were set to be $\mathcal{N}(0,1)$, an uninformative transformation of the quadratic limb-darkening parameters \citep[see][]{Kipping2013LD}.

We tried fits assuming circular and non-circular orbits and the results favour the non-circular orbit, which gave an eccentricity of \KKe\, and an argument of periapsis of \KKw\, degrees.

The final fits for the K2 and LCO lightcurves are shown in red in Figures \ref{fig:k2_transit} and \ref{fig:lco_transit}, and for the radial-velocities in Figure \ref{fig:rv_curve}. The priors and posterior values of the fitted parameters with \texttt{exonailer} are listed in Table \ref{tab:planetparams}. As can be seen, the photometric jitter is significant only for the K2 lightcurve; the LCO photometric jitter is consistent with zero. This is due to the fact that we decided to estimate the errors directly from the K2 photometry, while the extra jitter was added in quadrature to the LCO errorbars given by the photometric pipeline. For the radial-velocity jitter, it can be seen that the extra term for both instruments is also consistent with zero.

\subsection{Planet Parameters}
\label{sec:planetparams}

\begin{table*}
	\centering
	\caption{Parameters of \KKplanet}
	\label{tab:planetparams}
	\begin{tabular}{lllr}
		\hline
        \hline
		Parameter & Units & Value & Priors*\\
        \hline
        Period & days & \KKperiod & $\mathcal{N}(6.569,0.01)$ \\
        T$_0$ & days & \KKtc & $\mathcal{N}(2457588.28544,0.01)$ \\
        T$_{14}$ & hours & \KKptdur \\
        T$_{23}$ & hours & \KKptdurflat \\
        T$_{12=34}$ & hours & \KKptgress \\
        R$_{\rm P}$/R$_*$ & & \KKpratio & $\mathcal{U}(0.05,0.2)$ \\
        b & & \KKpb \\
        i & $\degree$ & \KKpinc & $\mathcal{U}(80,90)$ \\
        a$_*$ & AU & \KKpa & $\mathcal{U}(3.0,30.0)$ [R$_*$] \\
        K & \kms & \KKK & $\mathcal{N}(0.1,0.1)$ \\
        $\gamma_{\rm CORALIE}$ & \kms & \KKpgammaC & $\mathcal{N}(1.22,0.05)$ \\
        $\gamma_{\rm HARPS}$ & \kms & \KKpgammaH & $\mathcal{N}(1.24,0.05)$ \\
        CORALIE jitter & \kms & \KKpjitterC & $\mathcal{J}(0.0001,0.1)$ \\
        HARPS jitter & \kms & \KKpjitterH & $\mathcal{J}(0.0001,0.1)$ \\
        Incident flux <F> & $10^8$ erg s$^{-1}$ cm$^{-2}$ & \KKpicflux \\
        e & & \KKe & $\beta (0.867,3.03)$ \\
        $\omega$ & $\degree$ & \KKw & $\mathcal{U}(0.0,180.0)$ \\
        M$_{\rm P}$ & \mjup & \KKpmass \\
        R$_{\rm P}$ & \rjup & \KKpradius \\
        log g$_{\rm P}$ & dex (cgs) & \KKplogg \\
        $\rho_{\rm P}$ & $\rho_{\rm Jup}$ & \KKprho \\
        T$_{\rm eq.}$ & K & \KKpeqtemp \\
		\hline
		\hline
        \multicolumn{4}{l}{* $\mathcal{N}(\mu,\sigma)$ is a normal distribution with mean $\mu$ and standard deviation $\sigma$; $\mathcal{U}(a,b)$ is a uniform distribution between}\\
        \multicolumn{4}{l}{values \textit{a} and \textit{b}; $\mathcal{J}(a,b)$ is a Jeffrey's distribution with a lower limit of \textit{a} and \textit{b}; and $\beta (a,b)$ is a Beta distribution}\\
        \multicolumn{4}{l}{with parameters \textit{a} and \textit{b} as described by \citet{Kipping2013}.}
	\end{tabular}
\end{table*}

\texttt{exonailer} was able to determine various system parameters from the light curve transit shape: $a/R_*$, the semi-major axis-to-stellar radius ratio; \textit{$R_p/R_*$}, the ratio of planetary to stellar radius; \textit{$t_0$}, the time of the first observed transit; \textit{P}, the orbital period of the planet; and \textit{i}, the inclination of the planet's orbit. Additionally, from the radial velocity curves: \textit{e}, the eccentricity; \textit{$\omega$}, the periapsis argument; and \textit{K}, the radial velocity semi-amplitude of the star. Through a combination of these parameters and the already determined stellar mass and radius from \S~\ref{sec:starparam}, further properties of the planet can be determined using the equations as described in \citet{SeagerMallenOrnelas}. We measured the planetary mass to be \KKpmass\mjup\, with a radius of \KKpradius\rjup. This indicates a bulk density which is slightly less than that of Jupiter, \KKprho\rhojup. The planet has an incident flux of \KKpicflux$\times 10^8$ erg s$^{\rm -1}$ cm$^{\rm -2}$. The predicted equilibrium temperature is \KKpeqtemp K, with the assumption of a blackbody and an efficient transfer of energy from the day- to night-side. These are all listed in Tab.~\ref{tab:planetparams}.

\section{Discussion}
In this section we compare the properties of \KKplanet\ to the population of known hot Jupiters, and for this purpose we use the NASA Exoplanet Archive \footnote{\url{exoplanetarchive.ipac.caltech.edu}} \citep{Akeson_et_al_2013} as accessed on 2017 June 6.

Of the known warm and hot Jupiters discovered from \textit{K2}, \KKplanet\, currently has the third longest orbital period (see Tab.~\ref{tab:HJ}). This further demonstrates the ability of \textit{K2} to find longer period planets as was done by \textit{Kepler}.

\begin{table*}
	\centering
	\caption{\textit{K2} Discovered warm and hot Jupiters with precise measurements (20\%) on the masses and radii.}
	\label{tab:HJ}
	\begin{tabular}{ccccc}
		\hline
        \hline
		Planet & P$_{\rm orb}$ (d) & Mass (\mjup) & Radius (\rjup) & Reference\\
        \hline
        K2-29b & $3.2588321\pm0.0000019$ & $0.73\pm0.04$ & $1.19\pm0.02$ & \citet{K229} \\
        " & $3.2589263\pm0.0000015$ & $0.613^{+0.027}_{-0.026}$ & $1.000^{+0.071}_{-0.067}$ & \citet{K229andK230} \\
		\hline
        K2-30b & $4.098503\pm0.000011$ & $0.579^{+0.028}_{-0.027}$ & $1.039^{+0.050}_{-0.051}$ & \citet{K229andK230} \\
        " & $4.098513\pm0.000018$ & $0.625\pm0.030$ & $1.197\pm0.052$ & \citet{K230andK234} \\
        " & $4.09849^{+0.00002}_{-0.00002}$ & $0.589^{+0.023}_{-0.022}$ & $1.069^{+0.023}_{-0.019}$ & \citet{K230andK234brahm} \\
        \hline
        K2-31b & $1.257850\pm0.000002$ & $1.774\pm0.079$ & $0.71-1.41$ & \citet{K231} \\
        \hline
        K2-34b & $2.9956675^{+0.0000075}_{-0.000071}$ & $1.649\pm0.098$ & $1.217\pm0.053$ & \citet{K230andK234} \\
        " & $2.995654\pm0.000018$ & $1.773\pm0.086$ & $1.44\pm0.16$ & \citet{K234} \\
        " & $2.995629^{+0.000006}_{-0.000006}$ & $1.698^{+0.061}_{-0.050}$ & $1.377^{+0.14}_{-0.13}$ & \citet{K230andK234brahm} \\
        \hline
        K2-60b & $3.00265\pm0.00004$ & $0.426\pm0.037$ & $0.683\pm0.037$ & \citet{K260andK2107} \\
        \hline
        K2-97b & $8.4016\pm0.0015$ & $1.10\pm0.11$ & $1.31\pm0.11$ & \citet{K297} \\
        \hline
        K2-99b & $18.249\pm0.001$ & $0.97\pm0.09$ & $1.29\pm0.05$ & \citet{K299} \\
        \hline
        K2-107b & $3.31392\pm0.00002$ & $0.84\pm0.08$ & $1.44\pm0.15$ & \citet{K260andK2107} \\
        \hline
        \hline
	\end{tabular}
\end{table*}

To compare the ability to find longer period warm and hot Jupiters from the ground, a subset of all confirmed planets with masses greater than 0.2\mjup\, (and with other significant parameters -- such as eccentricity, planet density, and planet radius -- also measured) were split by their discovery `location' (i.e. \textit{Kepler}, \textit{K2}, other space missions and missions from the ground). Currently, \textit{K2} has only found 8 planets within this subset.

\begin{figure}
	\includegraphics[width=\columnwidth]{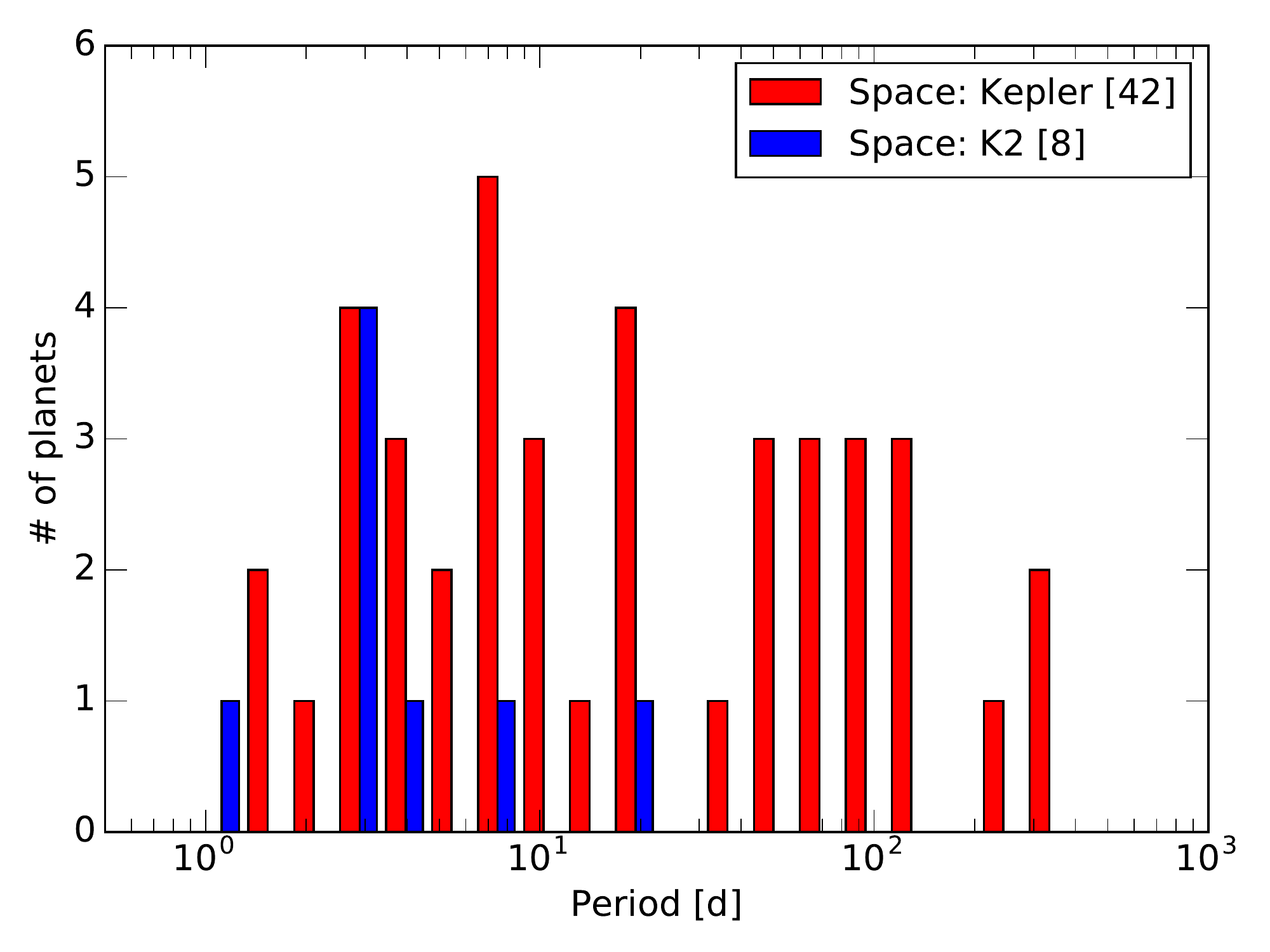}
    \caption{Distribution of confirmed planets found with \textit{Kepler} and \textit{K2} with masses $>\rm 0.2M_{\rm Jup}$. \textit{Kepler} is in red and \textit{K2} in blue. \KKplanet\, has been included into the \textit{K2} distribution.}
    \label{fig:histogram}
\end{figure}

The number of planets discovered by \textit{Kepler} and \textit{K2} over the period range strongly reveals that, as expected, \textit{K2} is less sensitive to long-period planets than \textit{Kepler}, equally sensitive to short-period planets (Fig. ~\ref{fig:histogram}). \textit{Kepler} observed 156,000 stars and, to date, \textit{K2} has observed a total of 171,610 (all stars observed by \textit{K2} in long cadence from Campaigns 1--10). However, there will be a natural cut off at $\sim$30-40d orbital period planets for \textit{K2} as campaigns typically do not last longer than 80 days, whereas \textit{Kepler} had almost 4 years of continuous observation of its stars. Additionally, due to the necessary follow-up time required per planet (radial velocity, imaging etc.) the community has had much longer to confirm \textit{Kepler} candidates compared with \textit{K2} candidates -- there are still regular announcements of discoveries from older \textit{K2} campaigns as well as discoveries from the current campaign. Given more time, the distribution for \textit{K2} may well fill out similarly to \textit{Kepler}.  By the conclusion of the \textit{K2} mission (assuming 19 full campaigns), we may expect \textit{K2} will produce more than double the number of transiting giant planets with periods <10\,d  compared to \textit{Kepler}.

\begin{figure}
	\includegraphics[width=\columnwidth]{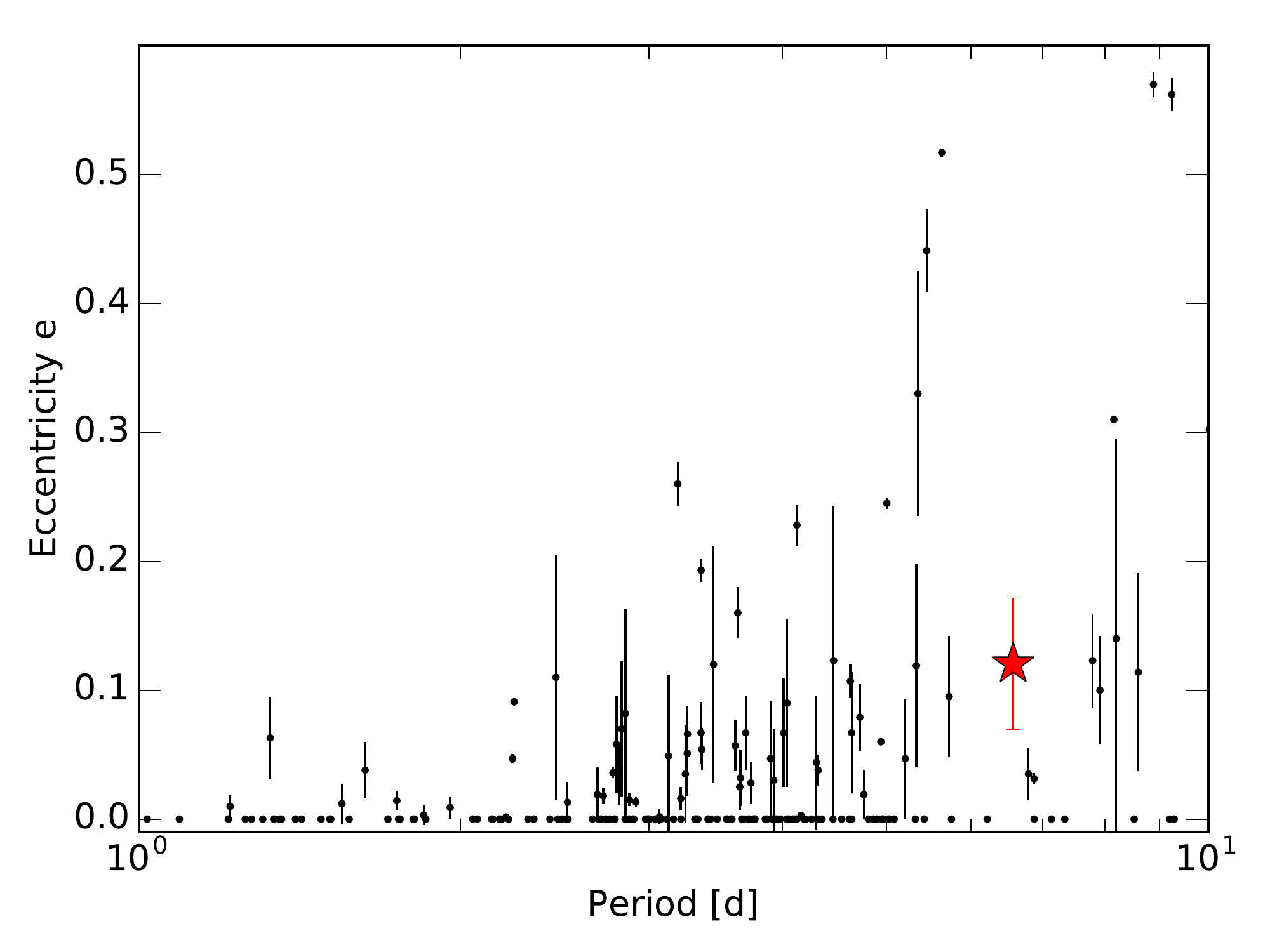}
    \caption{Eccentricities of transiting hot Jupiters (P=1--10\,d, \mplanet>0.2\mjup).  \KKplanet\ is plotted as a red star.  Planets with undetermined eccentricities have been excluded.}
    \label{fig:eccperiod}
\end{figure}

In Fig.~\ref{fig:eccperiod} we plot the measured eccentricities for transiting hot Jupiters with periods between 1--10\,d. Below $\sim5.5$\,d, approximately 70\% of hot Jupiters have measured eccentricity consistent with 0.  However for systems with period greater than 5.5\d, this fraction drops below 50\%.  It is therefore not surprising that we find a non-zero eccentricity for \KKplanet\ (e=\KKe).  If we assume a Q-factor of $10^6$ \citep{Wu2005} we calculate \citep{GoldreichSoter66} a tidal circularisation timescale of $\tau_{e}=$\KKte\ Gyr.  Given our best estimate for the age of the system (\KKage Gyr), this means that the timescale is of the same order of the age of the star.

\begin{figure}
	\includegraphics[width=\columnwidth]{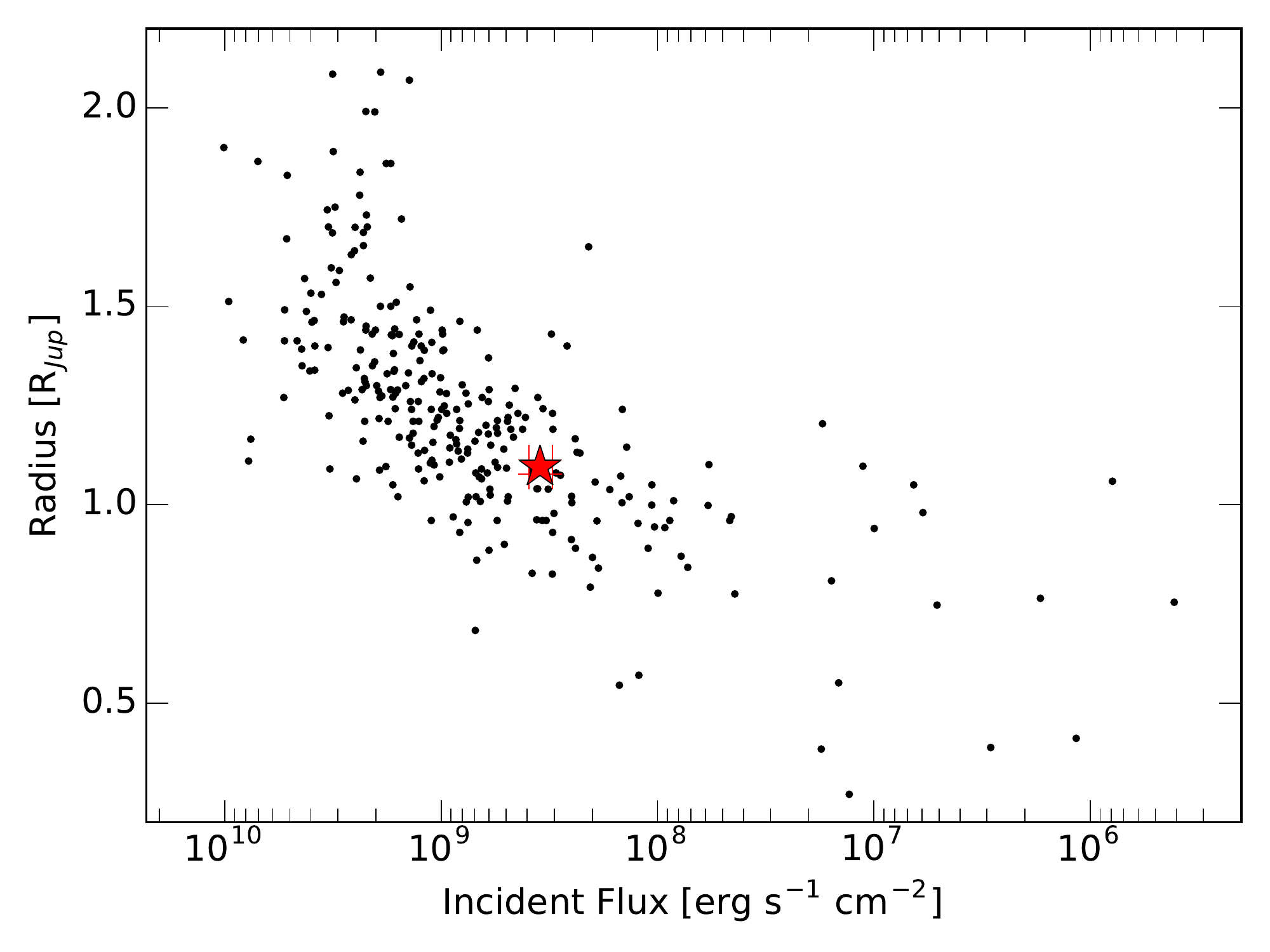}
    \caption{Distribution of incident flux and planet radii of confirmed planets found with measured masses (masses greater than $\rm 0.2M_{\rm Jup}$) and other measured properties. This work is represented by the red star and all other warm, hot Jupiters by black points.}
    \label{fig:radirrad}
\end{figure}

A common avenue of investigation associated with warm and hot Jupiters is determining whether they are inflated or not. Based on the  mass and radius of \KKplanet, it is slightly inflated compared with Jupiter, but not inflated with respect to other exoplanets with similar incident flux (see  Fig.~\ref{fig:radirrad}). The planet receives an incident flux of \KKpicflux$\times 10^8$ erg s$^{\rm -1}$ cm$^{\rm -2}$, which is very close to the empirical limit for inflation \citep[2$\times 10^8$ erg s$^{\rm -1}$ cm$^{\rm -2}$][]{DemorySeager11}.  Discovering exoplanet in this incident flux regime is important for studying the onset of the mechanism by which hot Jupiters are inflated.

\section{Conclusions}

We found a hot-Jupiter planet in data from \textit{K2} Campaign 10 and followed it up with radial velocity measurements and high angular resolution imaging. \KKplanet\, orbits a V=\KKVband, \KKage\,Gyr star with a [Fe/H] of \KKfeh. The planet has an non-circular orbit with an eccentricity of \KKe and period of \KKshortperiod\, days and a mass and radius of \KKpmass\mjup\, and \KKpradius\rjup\, respectively. It is the third longest period giant exoplanet discovered from \textit{K2} and has a period longer period than 94\% of giant planets discovered from ground-based transit surveys.

\section*{Acknowledgements}

We thank the Swiss National Science Foundation (SNSF) and the Geneva University for their continuous support to our planet search programs. This work has been in particular carried out in the frame of the National Centre for Competence in Research `PlanetS' supported by the Swiss National Science Foundation (SNSF).

N.E. acknowledges support from Financiamiento Basal PFB06. R.B., N.E., and A.J. acknowledge support from the Ministry for the Economy, Development and Tourism Programa Iniciativa Cient\'ifica Milenio through grant IC 120009, awarded to the Millenium Institute of Astrophysics.  A.J. acknowledges support by Fondecyt grant 1171208 and partial support by CATA-Basal (PB06, CONICYT). J.S.J. acknowledges support by Fondecyt grant 1161218 and partial support by CATA-Basal (PB06, CONICYT).

This work has made use of data from the European Space Agency (ESA)
mission {\it Gaia} (\url{https://www.cosmos.esa.int/gaia}), processed by
the {\it Gaia} Data Processing and Analysis Consortium (DPAC,
\url{https://www.cosmos.esa.int/web/gaia/dpac/consortium}). Funding
for the DPAC has been provided by national institutions, in particular
the institutions participating in the {\it Gaia} Multilateral Agreement.

This research was made possible through the use of the AAVSO Photometric All-Sky Survey (APASS), funded by the Robert Martin Ayers Sciences Fund.

This publication makes use of data products from the Two Micron All Sky Survey, which is a joint project of the University of Massachusetts and the Infrared Processing and Analysis Center/California Institute of Technology, funded by the National Aeronautics and Space Administration and the National Science Foundation.

Some/all of the data presented in this paper were obtained from the Mikulski Archive for Space Telescopes (MAST). STScI is operated by the Association of Universities for Research in Astronomy, Inc., under NASA contract NAS5-26555. Support for MAST for non-HST data is provided by the NASA Office of Space Science via grant NNX09AF08G and by other grants and contracts.

This paper includes data collected by the \textit{Kepler} mission. Funding for the \textit{Kepler} mission is provided by the NASA Science Mission directorate.

This research has made use of the NASA Exoplanet Archive, which is operated by the California Institute of Technology, under contract with the National Aeronautics and Space Administration under the Exoplanet Exploration Program.

This work makes use of observations from the LCO network.

The Robo-AO team thanks NSF and NOAO for making  the Kitt Peak 2.1-m telescope available. We thank the observatory staff at Kitt Peak for their efforts to assist Robo-AO KP operations. Robo-AO KP is a partnership between the California Institute of Technology, the University of Hawai`i, the University of North Carolina at Chapel Hill, the Inter-University Centre for Astronomy and Astrophysics (IUCAA) at Pune, India, and the National Central University, Taiwan. The Murty family feels very happy to have added a small value to this important project. Robo-AO KP is also supported by grants from the John Templeton Foundation and the Mt.~Cuba Astronomical Foundation. The Robo-AO instrument was developed with support from the National Science Foundation under grants AST-0906060, AST-0960343, and AST-1207891, IUCAA, the Mt.~Cuba Astronomical Foundation, and by a gift from Samuel Oschin. These data are based on observations at Kitt Peak National Observatory, National Optical Astronomy Observatory (NOAO Prop.~ID: 15B-3001), which is operated by the Association of Universities for Research in Astronomy (AURA) under cooperative agreement with the National Science Foundation. C.B. acknowledges support from the Alfred P.~Sloan Foundation. A.C.C. acknowledges
support from STFC consolidated grant number ST/M001296/1. DJA acknowledges support from STFC consolidated grant reference ST/P000495/1.

This research has made use of the SIMBAD database and of the VizieR catalogue access tool operated at CDS, France, and used the DACE platform developed in the frame of PlanetS (\url{https://dace.unige.ch}).



\bibliographystyle{mnras}
\bibliography{references}


\bsp	
\label{lastpage}
\end{document}